\documentclass[reprint,prl,twocolumn,showpacs]{revtex4-1}
\usepackage{txfonts}
\usepackage{float}
\usepackage{graphicx}
\begin{document}
\title{Origin of positive out-of-plane magnetoconductivity in overdoped Bi$_{1.6}$Pb$_{0.4}$Sr$_{2}$CaCu$_{1.96}$Fe$_{0.04}$O$_{8+\delta}$}
%\LaTeX\ Compuscript for \textit{Journal of the Physical Society of Japan}
\date{\today}
%\author{Tomohiro Usui$^1$\thanks{h12gs702@stu.hirosaki-u.ac.jp}, Shintaro Adachi$^1$, Takao Watanabe$^1$, Itsuhiro Kakeya$^2$, Akihiro Kondo$^3$, Koichi Kindo$^3$ and Shojiro Kimura$^4$}
\author{Takao Watanabe$^1$}
%\affiliation{Graduate School of Science and Technology, Hirosaki University, 3 Bunkyo, Hirosaki, 036-8561 Japan}
\author{Tomohiro Usui$^1$}
%\affiliation{Graduate School of Science and Technology, Hirosaki University, 3 Bunkyo, Hirosaki, 036-8561 Japan}
\author{Shintaro Adachi$^1$}
%\affiliation{Graduate School of Science and Technology, Hirosaki University, 3 Bunkyo, Hirosaki, 036-8561 Japan}
\author{Yuki Teramoto$^1$}
%\affiliation{Graduate School of Science and Technology, Hirosaki University, 3 Bunkyo, Hirosaki, 036-8561 Japan}
\author{Mihaly M. Dobroka$^1$}
%\affiliation{Graduate School of Science and Technology, Hirosaki University, 3 Bunkyo, Hirosaki, 036-8561 Japan}
\author{Itsuhiro Kakeya$^2$}
%\affiliation{Department of Electronic Science and Engineering, Kyoto University, Kyoto 615-8510, Japan}
\author{Akihiro Kondo$^3$}
%\affiliation{Institute for Solid State Physics, University of Tokyo, 5-1-5 Kashiwanoha, Kashiwa, Chiba 277-8581, Japan}
\author{Koichi Kindo$^3$}
%\affiliation{Institute for Solid State Physics, University of Tokyo, 5-1-5 Kashiwanoha, Kashiwa, Chiba 277-8581, Japan }
\author{Shojiro Kimura$^4$}
%\affiliation{Institute for Materials Research, Tohoku University, 2-1-1 Katahira, Aoba-ku, Sendai, 980-8577 Japan}
\affiliation{Graduate School of Science and Technology, Hirosaki University, 3 Bunkyo, Hirosaki, 036-8561 Japan$^1$}
\affiliation{Department of Electronic Science and Engineering, Kyoto University, Kyoto 615-8510, Japan$^2$}
\affiliation{Institute for Solid State Physics, University of Tokyo, 5-1-5 Kashiwanoha, Kashiwa, Chiba 277-8581, Japan$^3$}
\affiliation{Institute for Materials Research, Tohoku University, 2-1-1 Katahira, Aoba-ku, Sendai, 980-8577 Japan$^4$}
%\inst{$^1$Graduate School of Science and Technology, Hirosaki University, 3 Bunkyo, Hirosaki, 036-8561 Japan \\
%$^2$Department of Electronic Science and Engineering, Kyoto University, Kyoto 615-8510, Japan \\
%$^3$Institute for Solid State Physics, University of Tokyo, 5-1-5 Kashiwanoha, Kashiwa, Chiba 277-8581, Japan \\
%$^4$Institute for Materials Research, Tohoku University, 2-1-1 Katahira, Aoba-ku, Sendai, 980-8577 Japan} %\\
\begin{abstract}
To elucidate the pseudogap phase diagram including the overdoped state of high transition temperature (high-$T_c$) cuprates, we must understand the origin of the positive out-of-plane magnetoconductivity (MC) observed in these compounds. For this purpose, the out-of-plane resistivity $\rho_c(T,H)$ of an overdoped Bi$_{1.6}$Pb$_{0.4}$Sr$_{2}$CaCu$_{1.96}$Fe$_{0.04}$O$_{8+\delta}$ (Bi-2212) single crystal is measured under pulsed magnetic fields up to 60 T. We show that the superconducting density-of-states (DOS) depletion effect, in addition to the pseudogap effect, clearly appears below the superconducting fluctuation regime, and the contribution becomes dominant in the superconducting state.
%typically observed in high transition temperature (high-$T_c$) cupratesAbove the superconducting transition temperature $T_c$, the magnetoconductivity (MC) is due to two positive components: one component originates from the superconductive density-of-states (DOS) depletion effect, and the other component originates from the pseudogap effect. Subsequent analysis below $T_c$ shows that the positive slope for MC primarily originates from superconductivity.A simple extrapolation of the pseudogap contribution into the superconducting state accounts for a small percentage of the observed MC it value. This result implies that the pseudogap-like $\rho_{c}(T,H)$ peak structure observed in this overdoped Bi-2212 crystal primarily originates from superconductivity.
\end{abstract}
%To investigate the origin of the anomalous interlayer transport properties of high superconducting transition temperature Tc (high-Tc) cuprates,  under the high magnetic fields.
%for the negative magnetoresistance (MR), which is extracted by the data above the zero-field $T_c$, is found to
%%% Keywords are not needed any longer. %%%
%\kword{cuprate, superconductive fluctuation, interlayer conductivity, pseudogap effect, peak strusture  \ldots}
%%%
\maketitle
\section{I. Introduction}
%You can use this file as a template to prepare your manuscript for \textit{Journal of Physical Society of Japan} (JPSJ)\cite{jpsj,instructions}. No sections or appendices should be given to other categories than Regular Papers. Key words are not necessary.
To determine the mechanism of high-$T_c$ superconductivity, we must understand the relationship between the pseudogap and superconductivity~\cite{keimer}. To date, different classes of theoretical models for high-$T_c$ superconductivity have proposed different pseudogap phase diagrams~\cite{alloul}. That is, if the pseudogap is a necessary ingredient for pairing, the pseudogap opening temperature $T^{*}$ may merge with $T_c$ in the overdoped state, whereas if the pseudogap is of a competing order, $T^{*}$ may cross the $T_c$-``dome" near the optimal doping. The difference especially appears in the overdoped region; hence, it is important to investigate this region. However, it is difficult to determine whether the observed effect is related to the pseudogap or the superconductivity when $T^{*}$ is close to $T_c$.

It is well known that $\rho_{c}(T)$ shows a typical upturn~\cite{wata2} and a negative $\rho_{c}(H)$ slope~\cite{t.shiba,t.shiba1} at high fields below $T^{*}$. Because $\rho_{c}$ probes the electronic density-of-states (DOS) around the Fermi level, reflecting the tunneling nature between CuO$_2$ planes in high-$T_c$ cuprates, the semiconductive upturn in $\rho_{c}(T)$ is attributed to the decrease of DOS caused by the pseudogap opening, and the negative $\rho_{c}(H)$ slope is attributed to the DOS recovery along with the suppression of the pseudogap under magnetic fields. These facts are well established, and thus, we can usually estimate $T^{*}$ by the $\rho_{c}$ behavior.

On the other hand, because superconductivity is also a phenomenon connected to the energy gap, there have been proposals from both experimental~\cite{yur,lav,moro} and theoretical~\cite{gray,varla} points of view that superconductivity causes similar effects on $\rho_{c}$ in a temperature and magnetic field range in which the DOS effect dominates over the Cooper pair tunneling effect. In such cases, it is difficult to distinguish $T^{*}$ from the onset temperature of the superconducting fluctuation $T_{scf}$. However, this interpretation for the superconductivity-originated anomalous $\rho_{c}$ behavior has not been generally accepted. 

To address this issue, we focus on the magnetic field dependence of $\rho_{c}$ up to high fields, especially in the superconducting fluctuation regime. In general, the pseudogap is less sensitive to magnetic fields than superconductivity~\cite{kra}. Therefore, if we can obtain a high enough field near the upper critical field $H_{c2}$, we can expect to observe the pseudogap and superconducting contributions separately for the magnetic field dependence of $\rho_{c}$. Hence, we measure $\rho_{c}(T,H)$ of Fe-substituted overdoped Bi-2212 under high pulsed magnetic fields up to 60 T. For comparison, we also measure $\rho_{ab}(T,H)$ of the same sample.

\begin{figure*}[t]
\begin{center}
\includegraphics[width=180mm]{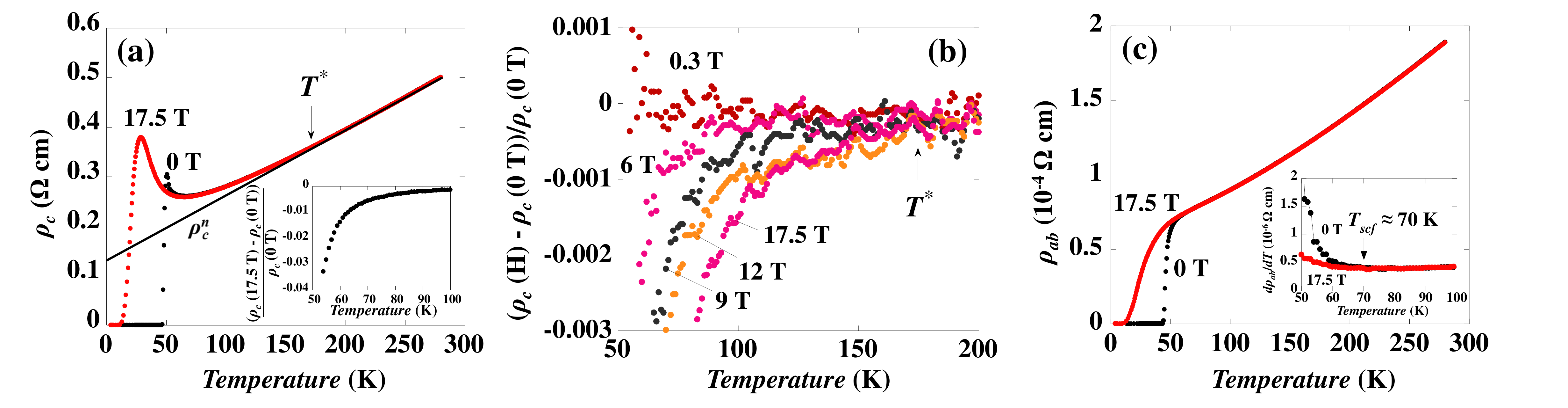}
\caption{\label{fig1}(Color online) (a) Out-of-plane resistivity $\rho_{c}(T)$ for
Bi$_{1.6}$Pb$_{0.4}$Sr$_{2}$CaCu$_{1.96}$Fe$_{0.04}$O$_{8+\delta}$ with and without a 17.5-T
magnetic field. Here, $\rho_{c}^{n}$ is a linear extrapolation of $\rho_{c}(T)$ at higher temperatures.
The arrow indicates the pseudogap opening temperature $T^* \approx$ 166 K.
The inset shows the negative magnetoresistance (MR) on $\rho_{c}$ under 17.5 T.
(b) Temperature dependence of the out-of-plane MR under several fields.
(c) In-plane resistivity $\rho_{ab}(T)$ with and without a 17.5-T magnetic field. The inset shows the temperature dependence of $d\rho_{ab}/dT$.
The arrow indicates the onset temperature of the superconducting fluctuation $T_{scf} \approx$ 70 K.}
%The inset shows the negative magnetoresistance (MR) on $\rho_{c}$ under 17.5 T.(b) Temperature dependence of MR under several fields.(c) Temperature dependence of $\rho_{ab}(T)$ with and without a 17.5-T magnetic field.\label{FIgure1.pdf}
\end{center}
\end{figure*}

\section{II. Experiment}
Single crystals were grown in air using the traveling-solvent floating zone (TSFZ) method.
The Bi site was partially substituted by Pb to overdope the sample. The Cu site was also
substituted by 2\% Fe to reduce $T_c$. The crystals were annealed under flowing oxygen at
400$^\circ$C for 50 h to promote hole doping ($T_c$ = 50 K, $p$ = 0.22). In this study, $T_c$ was determined by the onset of zero resistivity. The doping level
($p$) was obtained using the empirical relation~\cite{tallon} with maximum $T_c$ = 71 K for this sample.
%assumed to be the same as that of the non-Fe-substituted sample~\cite{usu} annealed under the similar condition, because it is known that $p$ did not significantly change with Fe substitution~\cite{maeda}.

%estimated using the empirical relation proposed by Tallon~\cite{tallon} for anon-Fe-substituted sample~\cite{usu}. Here, we assumed that $p$ was not significantly changed by
\begin{figure*}[t]
\begin{center}
\includegraphics[width=160mm]{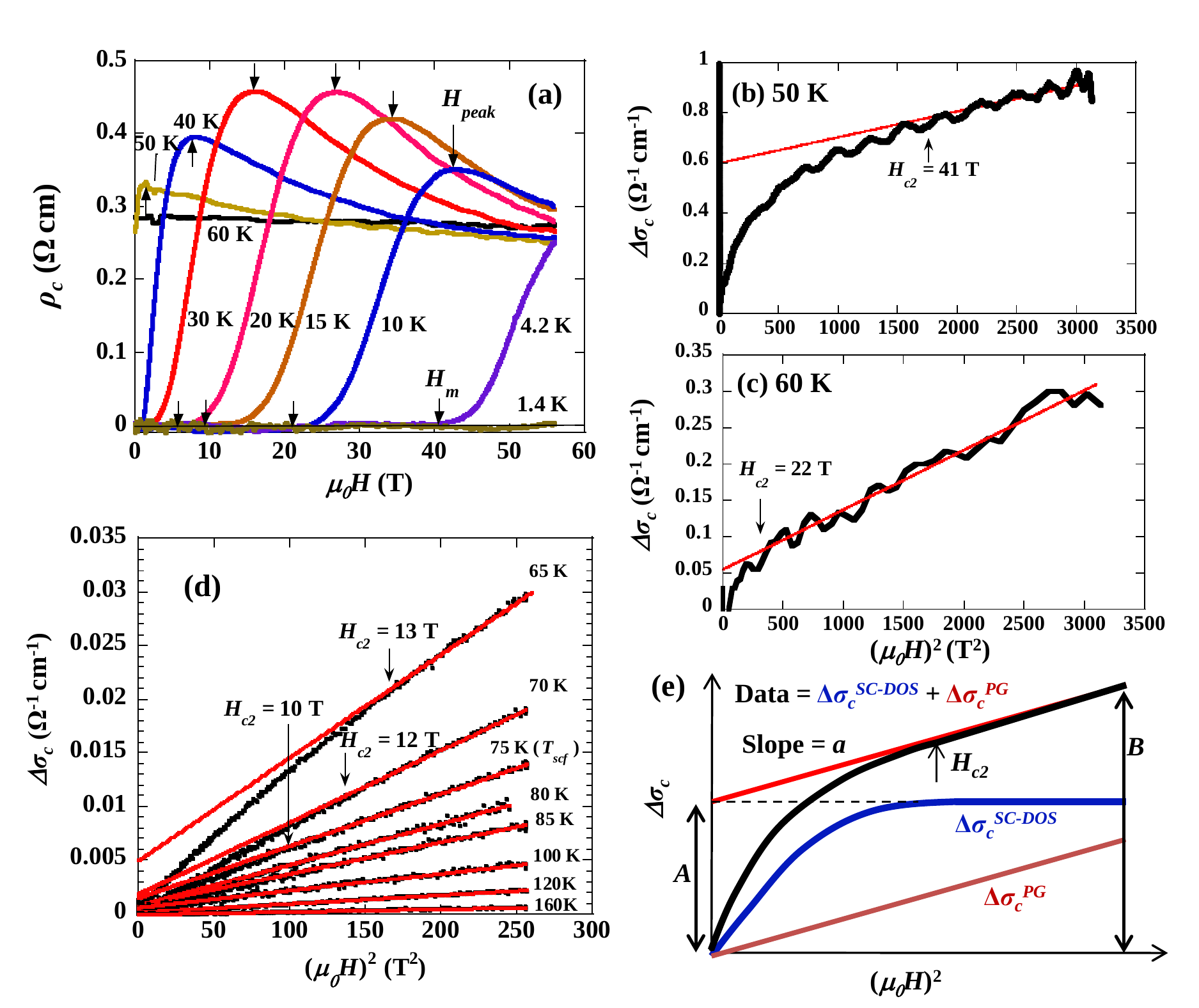}
\caption{\label{fig2}(Color online) (a) Magnetic field dependence of $\rho_{c}$ for
Bi$_{1.6}$Pb$_{0.4}$Sr$_{2}$CaCu$_{1.96}$Fe$_{0.04}$O$_{8+\delta}$ below and above $T_{c}$. The arrows
indicate the peak field $H_{peak}$, and the vortex melting field $H_{m}$ ($H_{m}$ is defined as a field in which $\rho_{c}$ = 0.01$\rho_{c}^{n}$~\cite{kadowaki}).
(b)--(c) Out-of-plane magnetoconductivity (MC), $\sigma_{c}(H)$--$\sigma_{c}(0$ T$)$, at 50 and 60 K, respectively. In (b), the normal state $\sigma_{c}(0$ T$)$ was assumed as an extrapolated value to the zero field of the high-field data.
(d) Out-of-plane MC above 65 K for steady magnetic fields. (b)--(d) The solid lines are linear extrapolations of MC at higher fields. Here, $H_{c2}$ is defined as the field at which MC deviates $1\%$ as an average from these lines. (e) Schematic representation for the two component analysis used above $T_{c}$.}
%Magnetic field dependence of $d\sigma_{c}/dH$ at $T_{c}$ = 50 K.
%Here, the log-term ($a\log(H/H_{c2})$) and linear-term ($bH$) were considered. The log-term was set to zero at $H \ge H_{c2}$.
%The parameters were selected as follows (in units of $\Omega^{-1}$cm$^{-1}$T$^{-1}$ for $a$ and $\Omega^{-1}$cm$^{-1}$T$^{-2}$ for $b$):
%(b) and (e) $a$ = -4.10$\times$10$^{-2}$, $H_{c2}$ = 41 T, and $b$ = 2.45$\times$10$^{-4}$,
%(c) $a$ = -1.20$\times$10$^{-2}$, $H_{c2}$ = 22 T, and $b$ = 1.74$\times$10$^{-4}$, and
%(d) $b$ = 1.45$\times$10$^{-4}$.
%, (e) $b$ = 4.19$\times$10$^{-5}$, and (f) $b$ = 5.17$\times$10$^{-5}$\label{f2}is the maximal field of the peak structure(the field at when $d\rho_{c}/dH$ = 0) and $H_{m}$ is the vortex melting field (the field at when$\rho_{c}$ = 0.01$\rho_{c}^{n}$, where $\rho_{c}^{n}$ is the bare resistivityextrapolated from high-temperature $T$-linear behavior
\end{center}
\end{figure*}

The in-plane resistivity $\rho_{ab}(T,H)$ as well as the $\rho_{c}(T,H)$ measurements were
performed using the DC four-terminal method~\cite{usu}. Two types of magnets were used:
the magnet at the Institute for Materials Research (IMR), Tohoku University, which
provides steady magnetic fields up to 17.5 T, and the nondestructive pulsed magnet at
the Institute for Solid State Physics (ISSP), University of Tokyo, which provides
pulsed magnetic fields (36-ms pulse duration) up to 60 T. Magnetic fields were applied
parallel to the c-axis. In the pulsed magnet measurements, we did not observe hysteresis behavior in high magnetic fields above the peak field $H_{peak}$ at which $\rho_{c}(H)$ is maximum. Furthermore, all down-ramped data coincided for measurements under several maximal fields, suggesting that the data were obtained with minimal eddy current influence.

In the following section, we will first estimate $T^{*}$ and $T_{scf}$ of this sample using the temperature dependence of $\rho_{c}$ and $\rho_{ab}$, respectively. Then, we will show that the out-of-plane magnetoconductivity (MC) is composed of two positive components below $T_{scf}$. This two component analysis enables us to estimate $H_{c2}$ at several temperatures. We will also estimate $H_{c2}$ from the magnetic field dependence of $\rho_{ab}$ in a standard manner. Consequently, an experimental $H$--$T$ phase diagram of this compound will be used to validate the analysis. Finally, we will discuss the implications of the obtained results for the pseudogap phase diagram.   
%We minimized the noise by two ways: (i) with making as small Au-wires-loops as possible against the magnetic fields to suppress electromagnetic induction, and (ii) with fixing Au-wires to the sample by covering all of them with apiezon-grease not to be vibrated by the strong Lorentz-force; the vibration causes a serious noise.For the $\rho_{c}(T,H)$ measurements, the voltagecontacts were attached to the center of the {\it ab} plane, and the current contactscovered almost all of the remaining space.Two types of magnets were used:the magnet at the Institute for Materials Research (IMR), Tohoku University, whichprovides steady magnetic fields up to 17.5 T, and the nondestructive pulsed magnet atthe Institute for Solid State Physics (ISSP), University of Tokyo, which providespulsed magnetic fields (36-ms pulse width) up to 60 T.show the magnetic field dependence of $\rho_{c}$, and
%\section{Results and Discussion}

\section{III. Results and discussion}
Figure \ref{fig1}(a) shows $\rho_{c}(T)$ for a
Bi$_{1.6}$Pb$_{0.4}$Sr$_{2}$CaCu$_{1.96}$Fe$_{0.04}$O$_{8+\delta}$ crystal with and without
a steady 17.5-T magnetic field. For zero field, $\rho_{c}(T)$ is metallic ($d\rho_{c}(T)/dT \textgreater$ 0)
over a wide temperature region, showing a slight upward trend with decreasing temperature.
From this result, we estimate the pseudogap opening temperature $T^*$ to be 166 K.
Here, $T^*$ is defined as the temperature at which $\rho_{c}$ increases by 1\% from that for
high-temperature linear behavior~\cite{wata1,wata2,usu}.
%First, we overview the normal state properties. the temperature dependence of

Figure \ref{fig1}(b) shows the
temperature dependence of the out-of-plane magnetoresistance (MR) for several magnetic fields $H$. A small negative MR is observed below 170 K, and the magnitude
increases with increasing magnetic field. This negative MR confirms that the pseudogap opens below this temperature~\cite{lav,t.shiba}. However, for a zero field, $\rho_{c}(T)$ shows a steep increase near $T_{c}$ (Fig. \ref{fig1}(a)). When a
17.5-T magnetic field is applied, $\rho_{c}(T)$ increases below $T_{c}$, showing
a sharp maximum around 30 K and then decreases to zero (Fig. \ref{fig1}(a)). The inset of
Fig. \ref{fig1}(a) shows an enlarged view of the out-of-plane MR at 17.5 T. The
small negative MR starting below $T^*$ rapidly increases near $T_{c}$. These results suggest that
the superconducting DOS fluctuation effect is added to the pseudogap effect below the onset temperature of the superconducting fluctuation $T_{scf}$~\cite{usu}.

To estimate $T_{scf}$, $\rho_{ab}(T)$ was measured with and without a steady 17.5-T magnetic field (Fig. \ref{fig1}(c)). The upward curvature for the temperature dependence is evident, further indicating that the sample is in an overdoped state~\cite{wata1,usu}. However, a small negative MR, whose origin is unknown, was observed above $T_{c}$ (details will be published elsewhere). Therefore, to estimate $T_{scf}$, we cannot use our standard method of determining the onset temperature for the rapid increase of positive MR~\cite{usu}. Instead, the temperature derivative $d\rho_{ab}/dT$ is compared for a zero field and a 17.5-T field (inset of Fig. \ref{fig1}(c)). The difference in $d\rho_{ab}/dT$ appears below 70--75 K. Because the Aslamazov-Larkin (AL)-type superconducting fluctuation effect (para-conductivity) is expected to affect the slope of $\rho_{ab}(T)$ below $T_{scf}$, the onset temperature for the change may be assigned as $T_{scf}$. Note that $T^*$ is far greater than $T_{scf}$, even in this overdoped state ($p$ = 0.22)~\cite{usu1}.
%To estimate $T_{scf}$, $\rho_{ab}(T)$ was differentiated with respect to the temperature. The inset shows the result.When the magnetic field is applied, we expect a $\rho_{ab}(T)$ slope smaller than the zero field slope below $T_{scf}$~\cite{usu1} because the positive Aslamazov-Larkin (AL) superconductive fluctuation contribution to conductivity is suppressed by the magnetic field.  Therefore,However, a small unknown negative MR, which may be caused by the Fe substitution, was observed in the normal state.

Figure \ref{fig2}(a) shows the high pulsed magnetic field data for $\rho_{c}(H)$ above and below $T_{c}$.
At 1.4 K, $\rho_{c}(H)$ remains at zero up to 55 T, indicating that $H_{c2}$ is above 55 T. At other temperatures below $T_{c}$, $\rho_{c}(H)$ shows a typical peak structure; it changes from zero to a resistive state at a vortex melting field, $H_{m}$, reaching a maximum at $H_{peak}$ before decreasing to a constant value. This peak structure in $\rho_{c}(H)$ can be understood by parallel two-channel tunneling conductivity~\cite{moro}: Cooper pair tunneling, $\sigma_{c}^{Cooper}(H)$, and quasiparticle tunneling, $\sigma_{c}^{qp}(H)$. That is, $\rho_{c}(H)$ = 1/($\sigma_{c}^{Cooper}(H)$ + $\sigma_{c}^{qp}(H)$). The behavior
under weak magnetic fields is controlled by $\sigma_{c}^{Cooper}(H)$, which arises due to phase slips between neighboring CuO$_{2}$ planes caused by the motion of mobile pancake vortices, and the Josephson contribution decreases with increasing field strength~\cite{Koshelev}.
 However, $\sigma_{c}^{qp}(H)$ increases under high magnetic fields, causing a negative $\rho_{c}(H)$ slope, and $H_{peak}$ is a crossover field for both contributions. In the following section, we concentrate on the analysis of the negative $\rho_{c}(H)$ slope. 
%the behavior of the high fields (i.e., )In the following section, we analyze the behavior of the high fields (i.e., the negative $\rho_{c}(H)$ slope). We identify two origins of the enhancement of $\sigma_{c}^{qp}(H)$ by applying magnetic fields: the suppression of (i) the pseudogap~\cite{t.shiba} and (ii) the superconducting gap~\cite{moro}; $\sigma_{c}^{qp}$ increases with increasing magnetic field because the depleted DOS owing to the opening of both gaps is partially recovered when the gaps are suppressed by magnetic fields.  Note that the negative $\rho_{c}(H)$ slope evolves smoothly even when the temperature is increased above $T_{c}$.
%the magnetic field dependence of$\sigma_{c}^{n}$ + When the temperature decreases, $H_{m}$ and $H_{peak}$ shift to the high field side.

We consider $\sigma_{c}(H)$ above $T_{c}$.
Figures \ref{fig2}(b)--(c) show the MC, $\Delta\sigma_{c}(H)$ = $\sigma_{c}(H) - \sigma_{c}(0)$, as a function of $H^2$, at 50 and 60 K, respectively. Above 65 K, we additionally measure the MC using a steady magnet up to 16 T (Fig. \ref{fig2}(d)).
The MC rapidly increases with increasing magnetic fields, approaching $\propto H^2$ for higher fields.
With increasing temperature, the field interval MC, which follows $\propto H^2$, extends to lower fields, and the slope, $a$, decreases. These data imply that the MC comprises two positive components: one component gradually increases with $aH^2$ (hereafter, we denote this component as $\Delta\sigma_{c}^{PG}$), and the other component rapidly increases with increasing magnetic field but tends to saturate at higher fields (hereafter, we denote this component as $\Delta\sigma_{c}^{SC-DOS}$). This two component analysis is schematically shown in Fig. \ref{fig2}(e). Figure \ref{fig2}(d) shows that the $H^2$ component, $\Delta\sigma_{c}^{PG}$, is present both below and above $T_{scf}$, while the component with a different (faster) $H$-dependence, $\Delta\sigma_{c}^{SC-DOS}$, decreases with increasing temperature and vanishes around $T_{scf}$ ( = 75 K). Thus, the former component ($\Delta\sigma_{c}^{PG}$) is attributed to the pseudogap effect, and the latter component ($\Delta\sigma_{c}^{SC-DOS}$) is attributed to the superconducting DOS depletion effect. This result verifies that the positive superconducting contribution in MC appears below $T_{scf}$. Furthermore, the observation of $H^2$ MC at high temperatures above $T_{scf}$ (when superconducting fluctuations are presumably gone) justifies the following analysis in Figs. \ref{fig2}(b)--(d), which separate the pseudogap components from its $H^2$ dependence.
%It should be noticeable that the the positive superconductive contribution (superconductive DOS fluctuation effect) in MC is successfully separated from the pseudogap contribution by using the data of magnetic field dependence up to high fields of 55 T. rapidly increases with increasing magnetic field but tends to saturate at higher fields , and the other component gradually increases with $aH^2$ .

This two component analysis enables us to estimate the weight of the superconducting contribution, $W_{SC}$, for the total quasiparticle tunneling MC. The saturated value of $\Delta\sigma_{c}^{SC-DOS}$ is given by the $y$-intercepts, $A$, of the high field $H^2$-linear behavior as shown in Fig. \ref{fig2}(e). The total quasiparticle contributions, $\Delta\sigma_{c}^{qp}$ = $\Delta\sigma_{c}^{SC-DOS} + \Delta\sigma_{c}^{PG}$, are given by $B$ of the figure. Then, we can estimate $W_{SC}$ = $\Delta\sigma_{c}^{SC-DOS}/(\Delta\sigma_{c}^{SC-DOS} + \Delta\sigma_{c}^{PG})$, by $A/B$. At 50 K, $W_{SC}$ is estimated as $66\%$ at 55 T. The estimated value is reliable in that it can be obtained directly from the raw MC data. Considering that the data are obtained at $T_{c}$, this value is fairly large. Furthermore, the field at which $\Delta\sigma_{c}^{SC-DOS}$ saturates is used to determine $H_{c2}$. Thus, $H_{c2}$ is estimated as 41, 22, 13, 12, and 10 T at 50, 60, 65, 70, and 75 K, respectively. However, $\Delta\sigma_{c}^{PG}$ does not saturate even under high fields up to 55 T. This indicates that the field needed to close the pseudogap, $H_{pg}$~\cite{t.shiba}, is very high (far greater than 55 T).
%at 70, 85, and 100 K are fitted and the results are shown in Fig. \ref{f2}(d), (e), and (f), respectively. In these cases, however, the log-term is absent.with the cut off field $H_{c2}$, as the data at 70 K under pulsed magnetic fields were noisy (Fig. \ref{fig2}(d))
%the magnetic field dependence of Here, the functional form of the log-term is a phenomenological one.
%$\sigma_{c}$ behavior does not able to explain only $H^2$. analyze this behavior in detail, Figure \ref{f2}(g) show the magnetic field dependence of $d\sigma_{c}/dH$ at 50 K.numerical fitting is performed with the simple two assumption that the one component is $a\log(H) + b$ (hereafter, we call this the log-term), and the other is $cH$ . because the positive values are unphysicalHere, we defined the field $H^{DOS}_{\sigma_{c}}$ at which the amplitude of log-term becomes zero. $H^{DOS}_{\sigma_{c}}$ is estimated as 40 and 22 T for 50 and 60 K.Moreover, we consider the integrated value of the log-term and linear-term, $\Delta\sigma_{c}^{log}$ = $\int_m^n[a log(H) + b]dH$ and $\Delta\sigma_{c}^{linear}$ = $\int_m^n[cH]dH$ ($m$ and $n$ are an initial and finite magnetic field.).This assumption representd a good fitting to experimental data in Fig. \ref{f2}(b)-(f). = $\Delta\sigma_{c}^{SC-DOS}/(\Delta\sigma_{c}^{SC-DOS} + \Delta\sigma_{c}^{PG})$ (Then, the total contribution for the former component to the MC is obtained from the y-intercepts of the figures.)to the total quasiparticle contributions at the highest field

\begin{figure*}[t]
\begin{center}
\includegraphics[width=140mm]{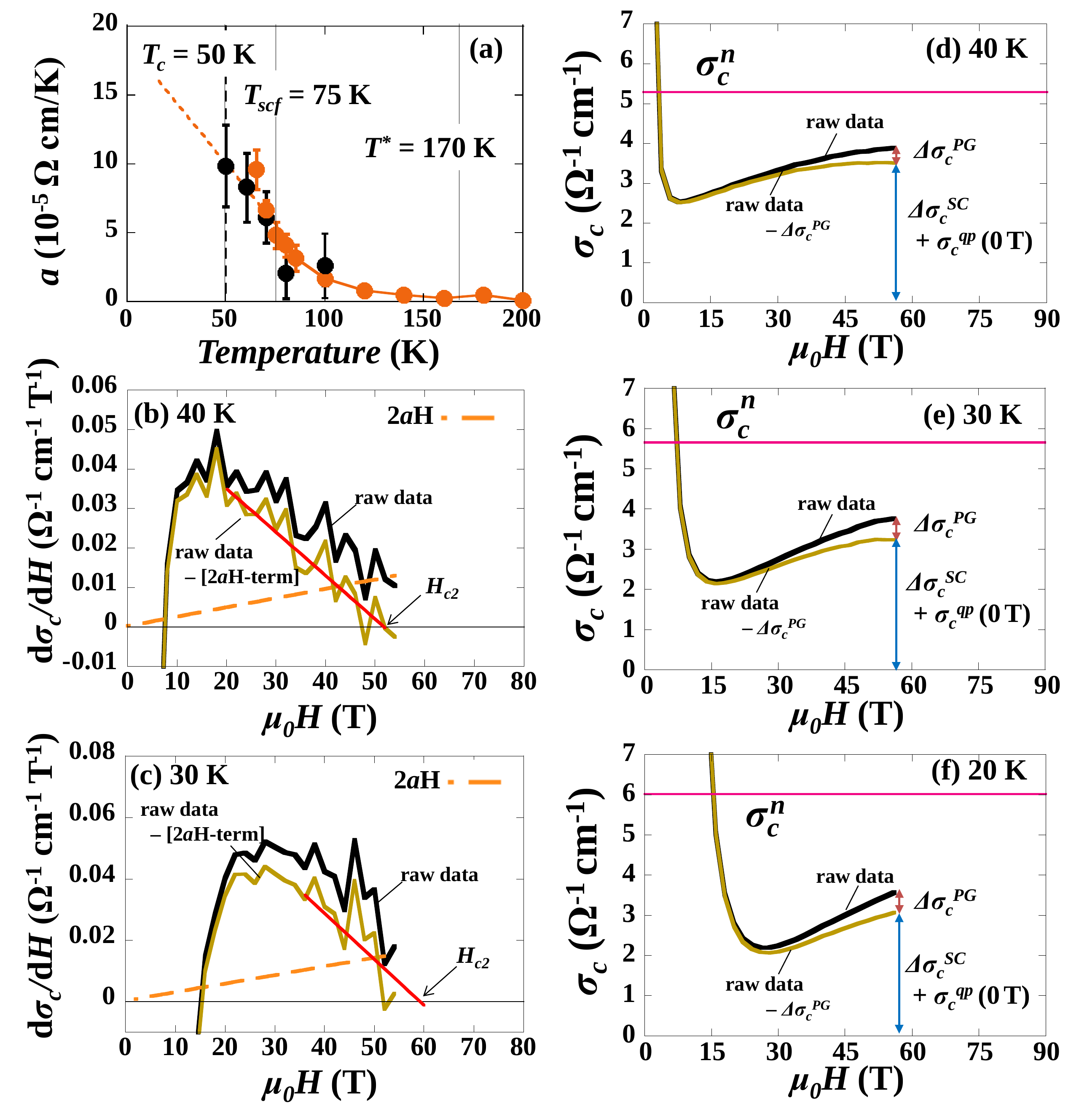}
\caption{\label{fig3}(Color online)
(a) Temperature dependence of $a$. Black and orange circles represent the estimation using data of a pulsed magnet (Figs. \ref{fig2}(b)--(c)) and of a steady magnet (Fig. \ref{fig2}(d)), respectively (for some data points, the raw MC data were not shown).
The dashed line represents a linear extrapolation of the $a$ values.
(b) and (c) Magnetic field dependences of the raw $d\sigma_{c}/dH$ data and the pseudogap contribution (2$aH$)-subtracted data for 40 K and 30 K, respectively.
The broken lines represent the pseudogap contribution. The solid lines represent linear extrapolations of the subtracted data.
%$H^{DOS}_{\sigma_{c}}$ is the field at which the extrapolation of the broken lines becomes zero.
(d)--(f) Magnetic field dependence of raw $\sigma_{c}$ data and the pseudogap contribution $\Delta\sigma_{c}^{PG}$-subtracted data at 40, 30, and 20 K, respectively.
Here, $\Delta\sigma_{c}^{SC-DOS}$ + $\sigma_{c}^{qp}(0$ T$)$ corresponds to the subtracted data.}
%\label{f3}
\end{center}
\end{figure*}

Next, we consider $\sigma_{c}(H)$ below $T_{c}$. Here, the $\sigma_{c}(H)$ behavior [positive $\sigma_{c}(H)$ slope] above $T_{c}$ is continuously observed above $H_{peak}$.
However, in this case, as MC does not exhibit $H^2$ behavior for whole fields, it is difficult to estimate the pseudogap contribution directly from the MC data. Therefore, we assume it from the data above $T_{c}$.
Figure \ref{fig3}(a) shows the $a$ values obtained in Figs. \ref{fig2}(b)--(d) for the pseudogap contribution above $T_{c}$. The accuracy of the data is not sufficient to determine the precise functional form (either sublinear or superlinear) for the temperature dependence of $a$. Thus, the $a$ value near $T_{c}$ was simply assumed to be linear in temperature. Note that this is an approximation near $T_{c}$ to estimate $a$ in the superconducting state. At higher temperatures ($\ge$ 80 K), the slope becomes smaller, and the $a$ value approaches zero at around $T^*$ = 170 K. 
Figure \ref{fig3}(b) and (c) show the magnetic field dependence of $d\sigma_{c}/dH$ at 40 and 30 K, respectively. To determine only the superconducting contribution to the data, the above pseudogap contribution is subtracted from the raw data. We find that the pseudogap contribution is not very large. In this case, $H_{c2}$ may correspond to the field at which the superconducting $d\sigma_{c}/dH$ is extrapolated to zero.
Thus, $H_{c2}$ is estimated as 51 and 60 T for 40 and 30 K, respectively.

We then investigate the weight of superconducting contribution $W_{SC}$ below $T_{c}$. For this purpose, the magnetic field dependences of $\sigma_{c}$ and $\sigma_{c}$ - $\Delta\sigma_{c}^{PG}$ are plotted in Figs. \ref{fig3}(d)--(f) at 40, 30, and 20 K, respectively. At the highest field (55 T), we consider $\sigma_{c}$ as $\sigma_{c}^{qp}$ because $\sigma_{c}^{Cooper}$ may be negligible. Then, $\sigma_{c}$ - $\Delta\sigma_{c}^{PG}$ is expressed as $\Delta\sigma_{c}^{SC-DOS}$ + $\sigma_{c}^{qp}(0$T$)$.
Assuming that $\sigma_{c}^{qp}(0$T$)$ at 20 K is negligible, the result in Fig. \ref{fig3}(f) implies that $W_{SC}$ accounts for $\approx$ 80 $\%$ of the total MC.
This indicates that the positive slope for the MC primarily originates from superconductivity.
%ithout $\Delta\sigma_{c}^{linear}$ and the large peak structure in $\rho_{c}(T)$Figure \ref{fig4} shows $\rho_{c}(T)$ for several magnetic fields depicted from the data shown in Fig. \ref{fig2}. We can recognize large peak structures below $T_{scf}$ in a similar manner as in ref. 5. However, the above result in $\sigma_{c}(H)$ indicates that these peak structures in $\rho_{c}(T)$ are also primarily originated from superconductivity. Indeed, the height of the peaks tends to decrease as the superconductivity is strongly destroyed under the high magnetic fields over 30 T.Figure \ref{fig2}(b) shows that, at 50 K, the superconducting contribution is 66$\%$ at 55 T. Considering that the data are obtained at $T_{c}$, this value is fairly large.

\begin{figure}[t]
\begin{center}
\includegraphics[width=80mm]{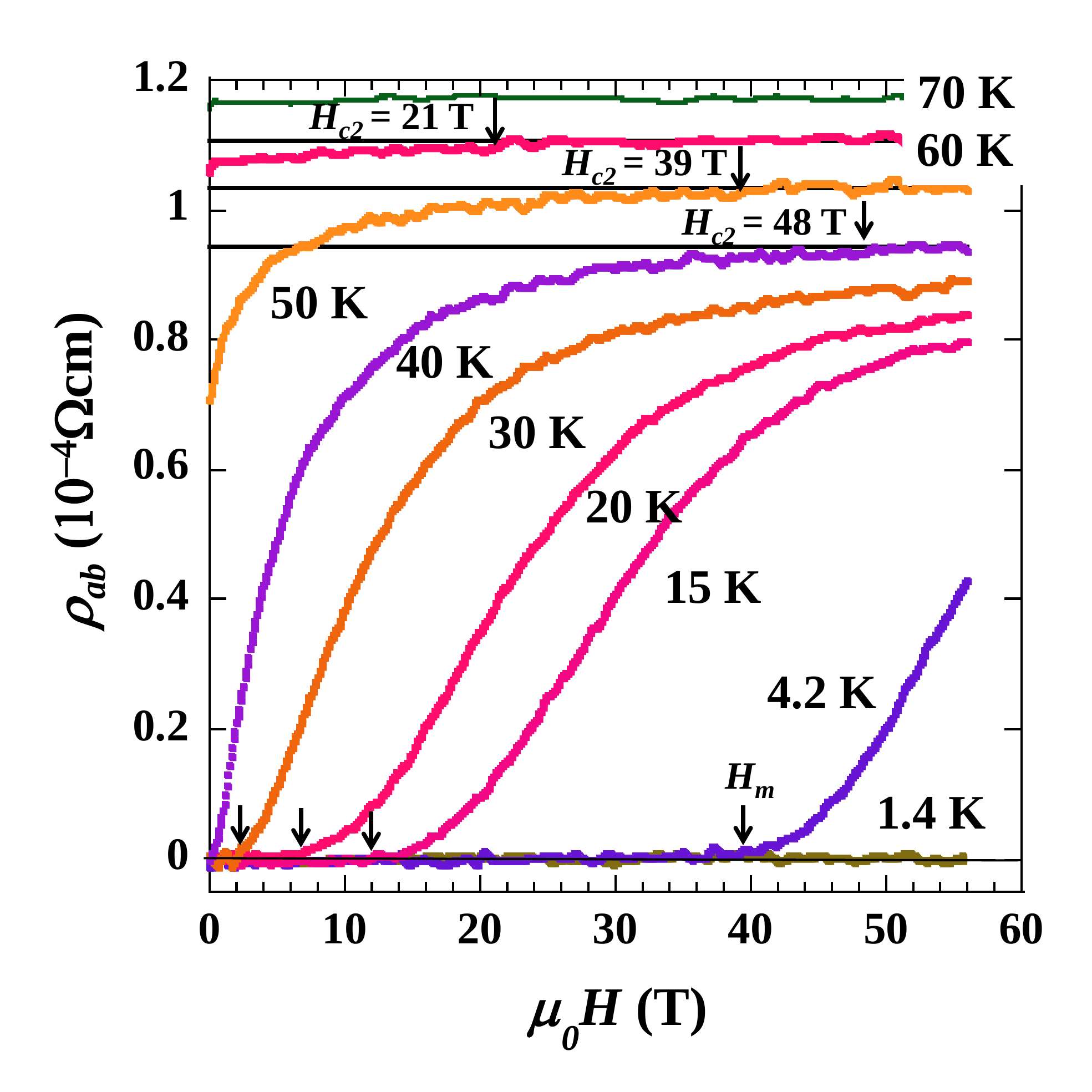}
\caption{\label{fig4}(Color online) Magnetic field dependence of $\rho_{ab}$ for
Bi$_{1.6}$Pb$_{0.4}$Sr$_{2}$CaCu$_{1.96}$Fe$_{0.04}$O$_{8+\delta}$ at several temperatures above and below $T_{c}$. The arrows
indicate the upper critical field $H_{c2}$ ($H_{c2}$ is defined as the field at which $\rho_{ab}$ decreases $1\%$ from the high field constant value), and the vortex melting field $H_{m}$ ($H_{m}$ is defined as a field in which $\rho_{ab}$ = 0.01$\rho_{ab}^{n}$).}
%$H_{m}$  shows the fitting result using the 2Dvortex solid state formula proposed by Blatter et al.\cite{blatter}.The dashed line for $H_{c2}$ is a guide for the eye.\label{f2}
\end{center}
\end{figure}

Figure \ref{fig4} shows the high pulsed magnetic field data for $\rho_{ab}(H)$ at several temperatures above and below $T_{c}$. With increasing magnetic fields, $\rho_{ab}(H)$ changes from zero to a resistive state at $H_{m}$ and rapidly increases in a similar manner as $\rho_{c}(H)$ (Fig. \ref{fig2}(a)). It then gradually approaches a constant value, keeping a positive slope. This in-plane behavior contrasts the out-of-plane behavior, in which $\rho_{c}(H)$ shows a typical negative slope at higher fields, indicating that, in Fig. \ref{fig2}(a), we have succeeded in detecting a property peculiar to $\rho_{c}(H)$. The superconducting fluctuation theory~\cite{varla} including the DOS contribution in highly anisotropic materials predicts that, for the out-of-plane conduction, the negative DOS contribution dominates over the positive Aslamazov-Larkin (AL) contribution near $H_{c2}$, resulting in a negative slope for $\rho_{c}(H)$. For the in-plane conduction, the AL contribution dominates, causing the positive slope for $\rho_{ab}(H)$. The experimental observations in this study are consistent with this theory. Then, it may be natural to ascribe the origin of the negative $\rho_{c}(H)$ slope (positive slope for the out-of-plane MC) primarily to the superconducting DOS fluctuation effect. Supposing that $\rho_{ab}(H)$ is constant above $H_{c2}$, $H_{c2}$ is estimated as 21, 39 and 48 T for 60, 50 and 40 K, respectively. These values agree with those obtained by $\rho_{c}(H)$ measurements, supporting the validity of the above analysis for $\rho_{c}(H)$.
%Thus, these $\rho_{ab}(H)$ results indicate that $H_{m}$ and $H_{c2}$ are correctly estimated by $\rho_{c}(H)$ measurements.indicates that the peak structure in $\rho_{c}(H)$ (Fig. \ref{fig2}(a))  The superconductive DOS fluctuation contribution could not be recognized in this in-plane data. The situation is the same as for our non-Fe-substituted sample~\cite{usu1}. The reason is an open question, as well as the results for the non-Fe-substituted sample~\cite{usu1}observed for higher fields (Fig. \ref{fig2}(a)) and this positive $\rho_{ab}(H)$ slope is consistent with the theory.  In addition, there is a theoretical proposal~\cite{mori} that, in the in-plane conduction of the $d$-wave superconductors such for high-$T_c$ cuprates, the positive regular Maki-Thompson (MT(reg)) contribution nearly cancels out the negative DOS contribution. This $d$-wave scenario may explain our observation (Fig. \ref{fig4}).However, in contrast to $\rho_{c}(H)$, $\rho_{ab}(H)$ Therefore, these results justify the above analysis that for both $\rho_{c}(H)$ and $\rho_{ab}(H)$ 

\begin{figure}[t]
\begin{center}
\includegraphics[width=80mm]{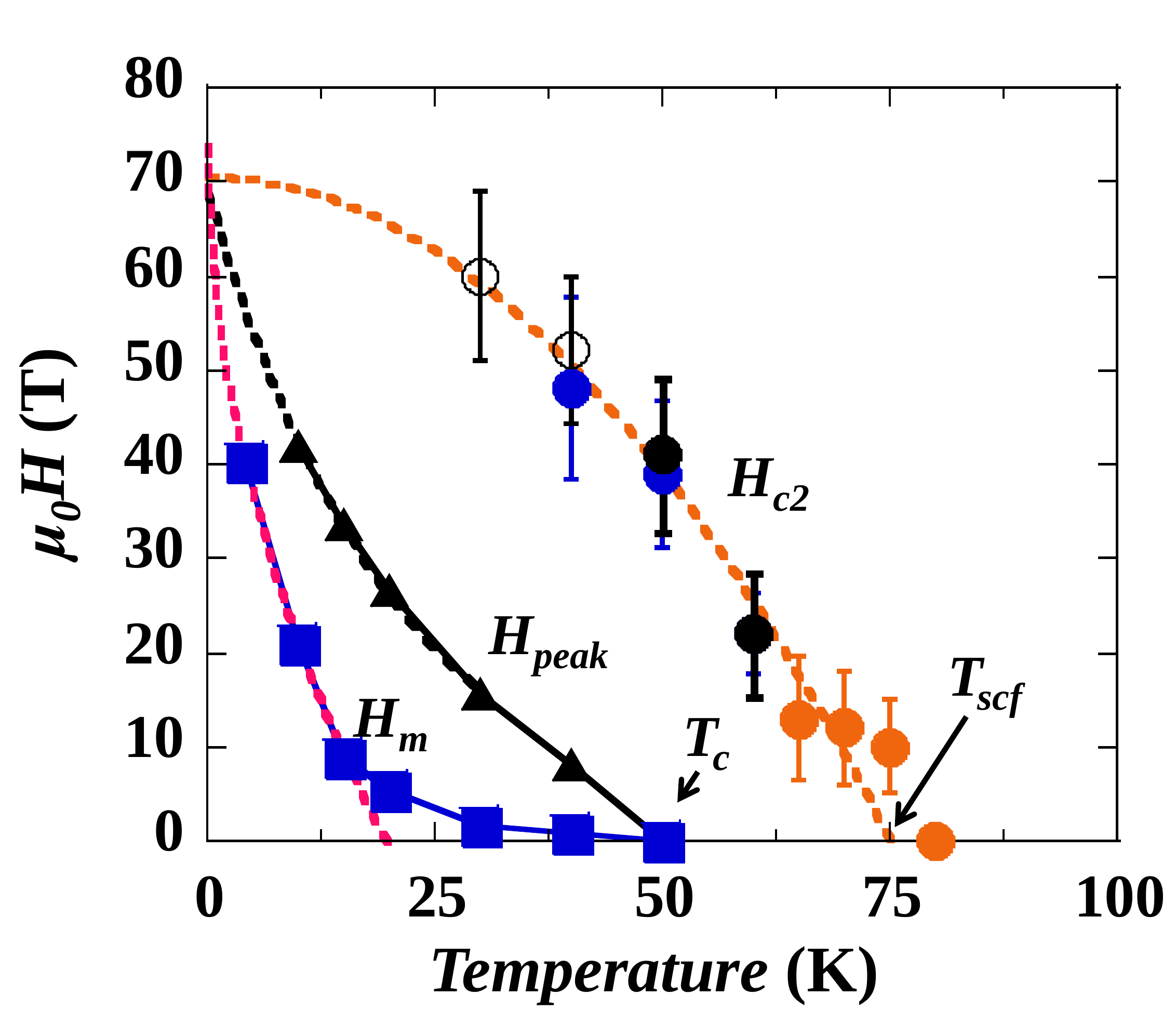}
\caption{\label{fig5}(Color online) Characteristic fields versus temperature. The melting field $H_{m}$,
peak field $H_{peak}$, and upper critical field $H_{c2}$ are shown as solid squares, solid triangles,
and circles, respectively. The $H_{c2}$ data points represented by black solid circles, blue solid circles, orange solid circles, and black open circles have been determined in Figs. \ref{fig2}(b)--(c), Fig. \ref{fig4}, Fig. \ref{fig2}(d), and Figs. \ref{fig3}(b)--(c), respectively. Here, an open symbol indicates that the data were obtained by extrapolation. The dashed lines are numerical fits for each field (see text).}
%$H_{m}$  shows the fitting result using the 2Dvortex solid state formula proposed by Blatter et al.\cite{blatter}.The dashed line for $H_{c2}$ is a guide for the eye.\label{f2}estimated with a pulsed magnet while orange symbols are estimated with a steady magnet. 
\end{center}
\end{figure}

Figure \ref{fig5} shows $H_{m}$, $H_{peak}$, and $H_{c2}$ obtained by $\rho_{c}(H)$ measurements as well as $H_{c2}$ obtained by $\rho_{ab}(H)$ measurements as a function of temperature.
In highly two-dimensional (2D) superconductors such as high-$T_{c}$ cuprates, the vortex system
transforms from a liquid state to a 2D solid state upon
cooling~\cite{blatter}. The transition temperature is represented as $T_{m}^{th}$. Here,
the field at which the vortex melting line extrapolates to 0~K is simply considered to be
$H_{c2}(0)$ at $T$ = 0. To estimate $H_{c2}(0)$, we adopted the
2D vortex solid state formula, $T_m^{th}/T_{c} = (c_{L}^2/Gi^{2D})(1-H/H_{c2}(0))^2$~\cite{blatter}. Here, we tentatively used the same parameter values, namely the Ginzburg number $Gi^{2D} = 0.05$ and the Lindemann number $c_{L} = 0.14$, as in the previous report~\cite{t.shiba1} with the exception of $T_{c}$ ( = 50 K). The fitting to the experimental data is very good between 4.2 K to 15 K, probably better than that of ref. ~\cite{t.shiba1}, and the fitting gives $H_{c2}(0)$ = 70 T.
The extrapolation of $H_{peak}$ to 0~K gives $H_{peak}$ = 69 T; here, we assumed $H_{peak}$ at
low temperatures is exponential in $T$~\cite{t.shiba}.
On the other hand, $H_{c2}(0)$ is directly estimated as 70 T by fitting the obtained $H_{c2}(T)$ using the empirical formula $H_{c2}(T) = H_{c2}(0)(1 - (T/T_{c})^{2})$; here, we assumed the $T_{scf}$ value ( = 75 K) for $T_{c}$. All three superconductivity-related characteristic fields roughly coincide at
$T$ = 0, indicating that the obtained $H_{c2}(T)$ contain information on the upper critical field. Hence, the above procedure (two component analysis in MC) for estimating $H_{c2}(T)$ from $\rho_{c}(H)$ data should be appropriate. In superconductors with strong fluctuation, the mean-field upper critical field, $H_{0}$, and the mean-field superconducting transition temperature in $H$ = 0, $T_{0}$, are shown to be the onset of superconducting fluctuation~\cite{ikeda}(see Fig. 3 in ref. ~\cite{ikeda}). We consider that $H_{c2}$ and $T_{scf}$ in this study correspond to $H_{0}$ and $T_{0}$, respectively.
%, proposed by Blatter et al.This result implies that the negative MR at the high field side, shown in Fig. \ref{f2}(a), primarily originates from superconductivity.For the estimation, we use the In particular, the Ginzburg number $Gi^{2D} = 0.05$ and the Lindemann number $c_{L} = 0.14$., and thus the above procedure for estimating $H_{c2}$ is justifiedThe coincidence between $H_{peak}$ at 0~K and $H_{c2}(0)$ suggests that $H_{peak}$ is an onset field of vortex liquid states., i. e., the attribution for positive $\sigma_{c}(H)$ slope at high fields to the destruction of the superconducting gap,

In a pioneering report~\cite{t.shiba1}, an unconventionally large quantum-dissipative pseudogapped state was shown to exist near $T$ = 0. However, we did not observe such behavior in our sample (Fig. \ref{fig5}). Note that $H_{m}$ is not observed until 55 T at 1.4 K (Fig. \ref{fig2}(a) and Fig. \ref{fig4}). We assume that our sample acquired a strong vortex pinning effect through the co-doping of Pb and Fe, and thus, such anomalous behavior has been suppressed compared to that in the pristine Bi-2212 sample of ref.~\cite{t.shiba1}.

Finally, we briefly discuss the implication of our result on the pseudogap problem. In this study, we have revealed that the superconducting contribution to the positive slope for the out-of-plane MC evolves below $T_{scf}$ in addition to the pseudogap contribution. This result, on one hand, confirms that the pseudogap and superconductivity are distinct, as many authors suggest~\cite{kra1,suzuki,tanaka}. On the other hand, this result verifies that superconductivity is another cause for the anomalous $\rho_{c}(T, H)$ behavior. Recently, we have observed a pseudogap-like upturn and negative MR in $\rho_{c}(T, H)$ just above $T_c$ for heavily overdoped Bi-2212 ($p$ = 0.23)~\cite{usu}. Thus, this behavior should be attributed to superconductivity; the pseudogap does not open above $T_c$ in this doping state. Therefore, if we could know whether or not the pseudogap opens below $T_{c}$ in this heavily overdoped sample, we would understand the pseudogap phase diagram completely.
\section{IV. Conclusions}

In summary, to determine the origin of the positive slope in interlayer MC in high-$T_c$ cuprates, $\rho_{c}(T,H)$ as well as $\rho_{ab}(T,H)$ measurements under high pulsed magnetic fields were performed on Pb and Fe co-doped Bi-2212. The results not only establish that the positive MC and the correlated upturn for $\rho_c(T)$ near $T_c$ are a sign of superconductivity but also enable us to estimate $H_{c2}$ and the superconducting contribution on the positive MC quantitatively. This finding may provide an important clue toward understanding the pseudogap phenomenon in high-$T_c$ superconductivity.
\section{Acknowledgments}
\begin{acknowledgments}

We thank T. Shibauchi, R. Ikeda, J. Goryo, and H. J. Im for their helpful discussions. The magnetoresistance measurements
were performed at the High Field Laboratory for Superconducting Materials, Institute for
Materials Research, Tohoku University (Project No.14H0007). The $\rho_{c}$ measurements under pulsed magnetic fields
were performed at the Institute for Solid State Physics, University of Tokyo. This work
is supported by JSPS KAKENHI Grant Number 25400349.
%For environments for acknowledgement(s) are available: \verb|acknowledgment|, \verb|acknowledgments|, \verb|acknowledgement|, and \verb|acknowledgements|.
\end{acknowledgments}
%\begin{figure}
% \begin{center}
%%\includegraphics{fig01.eps}
%\includegraphics[width=80mm]{Figure5-(usui)}
%\caption{(Color online) Temperature dependence of $\rho_{c}(T)$ under several magnetic fields,
%where $\rho_{c}^{n}$ is the bare resistivity estimated by the extrapolation from high temperature
%$T$-linear behavior. The open squares represent the resistivity under $H^{DOS}_{\rho_{c}}$, which
%is estimated by $\rho_{c}(T,H^{DOS}_{\rho_{c}}) = \rho_{c}(T,55 T) + \int_{55 T}^{H^{DOS}_{\rho_{c}}} %(a\log(H) + b)dH$.}
%\label{f5}
% \end{center}
%\end{figure}
%\appendix
%\section{}
%Use the \verb|\appendix| command if you need an appendix(es). The \verb|\section| command should follow even though there is no title for the appendix (see above in the source of this file).
%For authors of Invited Review Papers, the \verb|profile| command is prepared for the author(s)' profile.  A simple example is shown below.
%\begin{verbatim}
%\profile{Taro Butsuri}{was born in Tokyo, Japan in 1965. ...}
%\end{verbatim}

\end{document}